\documentclass[aps,prb,twocolumn,groupedaddress,floatfix,showpacs]{revtex4}

\usepackage{graphicx}

\newcommand{\be}{\begin{equation}}
\newcommand{\ee}{\end{equation}}
\newcommand{\bea}{\begin{eqnarray}}
\newcommand{\eea}{\end{eqnarray}}

\begin{document}


\title{Incomplete electronic relaxation and population upconversion
 in  quantum dots}


\author{Karel Kr{\'a}l}

\email[ ]{kral@fzu.cz}
\thanks{corresponding author}
\author{Zden{\v ek} Kh\'as}
\author{Petr Zden\v ek}
\affiliation{Institute of Physics, Academy of Sciences of Czech
Republic, Na Slovance 2, 18221 Praha 8, Czech Republic}


\date{\today}

\begin{abstract}
We calculate the  electronic relaxation  of a single electron in a
quantum dot  with two electronic orbital states,  and with the
electronic coupling to the longitudinal optical modes of the lattice
vibrations included in the  multiple scattering approximation to the
electronic self-energy. This model  shows that there is an incomplete
electronic relaxation from the excited state to the ground state and
also the incomplete increase of the population of the excited state
after resonantly exciting the ground state. These theoretical results
are compared with very similar recent findings reported in experimental
papers.
\end{abstract}

\pacs{73.,73.21.La,73.63.Kv,78.67.Hc}

\maketitle

\section{Introduction}
Relaxation of the electronic energy in individual quantum dots and the
question of the phonon bottleneck still remain to be of considerable
current interest
\cite{yoffe,bockelmann,benisty}. The Fr\"ohlich's mechanism of the electron-LO-phonon
coupling in the polar semiconductors is regarded as very important
\cite{ledentsov1998}. The electronic self-energy due to this
electronic interaction was studied by Inoshita and
Sakaki~\cite{inoshitasakakiSCB}, who found that the spectral features
of the electronic spectral density have very narrow width. The
dependence of these very narrow spectral peaks on the energy parameter
$E$  was shown to have the shape of the $E^{-1/2}$, in the limit of
zero temperature \cite{japrb5}.

The relaxation of the electron energy in quantum dots was studied later
\cite{japss1997,japss1998,tsuchiya,jaarXiv1} in the approximation of multiple
electronic scattering on optical phonons expressed in the
self-consistent Born approximation. It was theoretically shown that
despite the fact that the spectral peaks of the electronic spectral
density have zero width in the sense of full width at half maximum, the
relaxation rate of an electron from the excited state in the quantum
dot remains nonzero in the limit of zero temperature of the lattice
\cite{japss1997,tsuchiya}. The relaxation rate obtained in this way
was shown to have the  order of magnitude of 1/ps
\cite{japss1998,tsuchiya}, in agreement with broad
experimental data.

In the same approximation to the electronic self-energy it was
demonstrated that the model gives the electronic relaxation even in
cases when the electronic inter-level energy separation is not equal to
the optical phonon enery. The dependence of the relaxation rate on the
detuning, between the electronic energy level separation and the energy
of the optical phonon, has a resonance structure
\cite{jerusalem1998}.  The existence of this resonance
 structure is in accord with the
well-known experimental rule for detecting the presence of  quantum
dots in quantum dot samples with help of optical methods
\cite{ledentsov1998}, basing on the detection of the
resonance structure of the photoluminescence excitation signal with
peaks separated by  optical phonon energy $E_{LO}$.

 The important role
of  Fr\"{o}hlich's coupling has also been  supported by the  recent
interpretation of  the optical line  broadening and of the continuous
background observed in photoluminescence excitation spectra measured on
self-assembled quantum dot samples at high levels of the electronic
excitation
\cite{jaarXiv2}.

Treating the electronic relaxation in the self-consistent Born
approximation to the self-energy means that the electron makes the
irreversible transition between two electronic states which are
renormalized with respect to the electron-phonon coupling, with the
electronic self-energy taken in  the self-consistent Born
approximation. This approximation takes into account multiple
scattering processes of electron on the system of optical phonons and
introduces in this way the multiphonon states
 into the relaxation rate formula and into the formula for
the self-energy. The coherent multiphonon states created in this way
are analogous to  the classical macroscopic oscillations
\cite{terzi} of the lattice and the irreversible process of the electronic
relaxation in this system can be viewed as a quantum transition in a
potential dependent on time
\cite{taiwan}.  A nonconservation of energy in such a system then may be
 manifested as an absence of the   phonon
bottleneck in quantum dots. Because of the rapidity of the electronic
relaxation in quantum dots, this property  of the electronic
relaxation between the electronic states in the dot restricts
considerably the possible use of quantum dots  in constructing
quantum bits using  the electronic orbital states in quantum dots as
qubit states
\cite{tanamoto,wu}.

To our knowledge, in the theory the electronic relaxation has so far
been studied with an emphasis to the relaxation rate from the excited
state only, calculated at the innitial stage of the relaxation, which
was understood to be such instant of time, at which the electron is
created at this state
\cite{tsuchiya,japss1998}. The time dependence of the
electronic relaxation in quantum dots has not yet been given much
attention in theory. In this paper we present the numerical evaluation
of the time dependence of the relaxation. We  point out several
features which are seen from the numerical data, concerning the final
stage of the relaxation. In particular, we  demonstrate theoretically
the effect of the rapid decrease of the population of the ground state
energy level in the quantum dot and an appearence of electronic
population  in the excited state, after resonantly exciting the
electron to the ground state (an upconversion). The effect of the
incomplete depopulation of the electronic excited state in quantum dot,
at the end of the electronic relaxation from the higher-energy state
and at low temperatures, is also shown.  We point out  possible
connection of this theoretical effect to the recent experiments
\cite{urayama2001,quochi,urayama2002}.

\section{\label{}The transport equation}
 Similarly, as it was done in the previous
works, we  restrict ourselves to a simple model of the electronic and
phonon structure of the quantum dot.  First of all, we  ignore
completely the presence of the holes in the valence band states of the
dot. Although
  the holes may be simply assumed as being in thermal
  equilibrium with the surrounding
\cite{jiang},  we simplify our model assuming the holes to be heavy enough,
so that they do not  influence considerably the motion of the system
under consideration and contribute only to the static potential of the
quantum dot.

We consider the presence of a single electron only. The quantum dot
electronic orbital states, and the corresponding energies, are
approximated by the eigenstates of the infinitely deep cubic quantum
dot, with the lateral size $d$ and with the material properties of GaAs
crystal. We consider again the two-energy level model
\cite{jaarXiv1}, assuming that the electronic ground state
$\psi_0({\bf r})$, with the index $n=0$ and with the energy $E_0$, is
given by the electronic ground state in the infinitly deep cubic
quantum dot. The excited state, with $\psi_1({\bf r})$ being the wave
function of the state $n=1$ with the energy $E_1$, is chosen to be one
of the triple degenerate lowest energy excited states in this cubic
dot. We put $E_0=0$ in this work.

The electron is assumed to interact with the bulk modes of the
longitudinal optical phonons of lattice vibrations, neglecting in this
way any influence which the interfaces in the quantum dot may have on
the structure of the optical phonons. It was shown earlier that the
details of the optical phonons can be neglected in heterostructures
with not too small characteristic dimensions
\cite{ruckerapologet}. We confine ourselves to this simple model,
neglecting all other interaction, like the electronic coupling to the
acoustic phonons, and other electronic interactions. Neglecting the
electronic spin, the Hamiltonian of the electron-phonon system under
consideration reads~\cite{jaarXiv1,kkjaszowiec}
\be
H=H_0+H_1,
\ee
where $H_0$ is Hamiltonian of free electrons and free optical phonons,
\be
H_0=\sum_{n=0,1}E_{n}c^+_nc_n+\sum_{\bf q}b^+_{\bf q}b_{\bf q},
\ee
$c_n$ is annihilation operator of electron in state $n$, $b_{\bf q}$
is annihilation operator of optical phonon with wave vector ${\bf q}$.
The electron-phonon interaction operator (Fr\"ohlich's coupling) is
\be
H_1=\sum_{m,n=0,1} \sum_{\bf q} A_q\Phi(n,m,{\bf q})(b_{\bf q}-
b^+_{-{\bf q}})c^+_nc_m,
\ee
where the summation over ${\bf q}$ extends over the phonon wavevectors
in the first Brillouine zone. In the latter operator the coupling
constant is $A_q=(-ie/q)[E_{LO} (\kappa^{-1}_{\infty}-
\kappa^{-1}_0)]^{1/2}(2\varepsilon_0V)^{-1/2}$, where
$\kappa_{\infty}$ and $\kappa_0$ are, respectively, high-frequency and
static dielectric constants, $\varepsilon_0$ is permittivity of free
space, $-e$ is the electronic charge, ${q=\mid {\bf q}
\mid}$, and $V$ is volume of the sample.
$\Phi$ is the form-factor,
\be
\Phi(n,m,{\bf q})=\int d^3{\bf r}\psi^*_n({\bf r})e^{i{\bf qr}}\psi_m({\bf r}),
\ee
which modifies the original bulk form \cite{callaway} of the
electron-phonon operator to the case of quantum dot.

The relaxation rate $dN_1/dt$ can be obtained using the nonequilibrium
Green's functions method \cite{LL10}. Assuming the self-consistent
Born approximation for the electronic self-energy, the instantaneous
collisions approximation, the diagonal approximation for the
single-particle Green's functions and distribution functions, and the
Kadanoff-Baym ansatz \cite{kadanoff-baym} with the use of the
equilibrium spectral densities for the interacting system, we finally
get:
\begin{widetext}
\begin{eqnarray}
\frac{dN_1}{dt}=-\frac{2\pi}{\hbar} \alpha_{01}
\left[ N_1(1-N_0) \left( (1+\nu_{LO})  \nonumber
\int^{\infty}_{-\infty}dE\,\sigma_1(E)\sigma_0(E-E_{LO})
\label{rate}
+\nu_{LO}\int^{\infty}_{-\infty}dE\sigma_1(E)\sigma_0(E+E_{LO})
\right)  \right. \nonumber\\
- N_0(1-N_1)\left((1+\nu_{LO})
\int^{\infty}_{-\infty}dE\sigma_0(E)\sigma_1(E-E_{LO})
 \left.
+\nu_{LO}\int^{\infty}_{-\infty}dE \sigma_0(E)
\sigma_1(E+E_{LO})\right)\right]
\end{eqnarray}
\end{widetext}
We do not give here the details of the derivation of the relaxation
rate formula, referring the reader to
references\cite{jaarXiv1,tsuchiya} In formula (\ref{rate})
$\sigma_n(E)$ is spectral density of the state $n$ calculated in the
self-consistent Born approximation, $E$ being the energy variable.
$\nu_{LO}$ is Bose-Einstein distribution, giving the population of the
optical phonon modes at the given temperature $T$ of the lattice. The
constant $\alpha_{mn}$ is given by the definition
\be
\alpha_{mn}=\sum_{\bf q} \mid A_q \mid^2 \mid \Phi(n,m,{\bf
q})\mid^2, \qquad \alpha_{mn}=\alpha_{nm}.
\ee

The spectral densities, obeying the condition $\int_{-\infty}^{\infty}
\sigma_n(E)dE = 1$, are determined from the corresponding Green's
functions. The retarded electronic self-energy $M_n(E)$ of the state
$n$ is obtained, in the self-consistent Born approximation, from the
self-consistent equation
\begin{eqnarray}
M_n(E)=\sum_{m}\alpha_{nm}  \hspace{5cm} \nonumber \\ \times
\left( \frac{1-N_m+\nu_{LO}}{E-E_m-E_{LO}-M_m(E-E_{LO})+i0_+} \qquad
\right. \nonumber \\ \left.
+\frac{N_m+\nu_{LO}}{E-E_m+E_{LO}-M_m(E+E_{LO})+i0_+}.
 \right) \quad
 \label{SCB}
\end{eqnarray}

In the formula (\ref{rate}) for the relaxation rate the spectral
densities depend on the electronic density matrix by means of the
equation (\ref{SCB}) for the self-energy. This dependence of the
relaxation rate on the state of the electronic system is not easily
seen in a quantitative manner from the formula (\ref{rate}) and will be
therefore demonstrated numerically below. Let us remark, that  in the
case of approximating the electronic spectral densities   by
delta-functions, which would be a suitable approximation in the case of
very weak electron-phonon interaction, the integrals on the right hand
side of (\ref{rate}) would be performed and the dependence of the
relaxation rate on the population of the unperturbed electronic states
would be be simple. This approximation would give nonzero relaxation
rate only when the inter-level energy separation equals the optical
phonon energy. We shall therefore not pay any further attention to this
simple approximation.

\section{\label{}Electronic relaxation}

\begin{figure}[t]
\includegraphics[width=80mm]{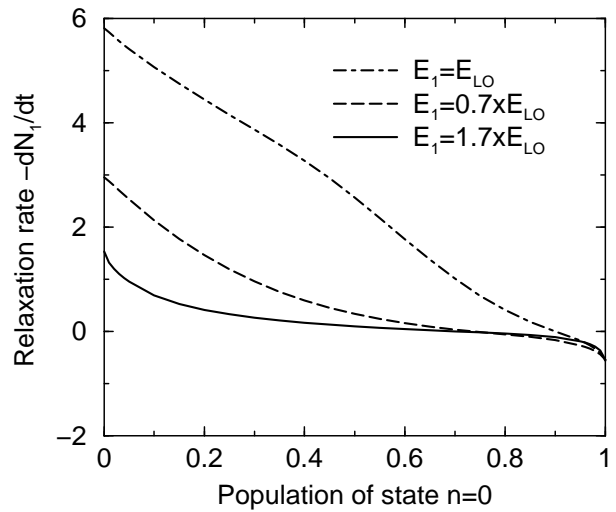}%
 \caption{Electronic relaxation rate $-dN_1/dt$ plotted for
 three values of the detuning, $\xi=1,\,\,\, 0.7$ and 1.7,
 as a function of the population
 $N_0$ of the lower-energy state $n=0$. Lattice temperature is $T$=10 K.}
 \label{tri}
\end{figure}

In the present approximation, in which we confine ourselves to the
diagonal terms of the electronic density matrix, given by $N_n$,
$n=0,1$, the electronic density matrix is determined by one of the
terms $N_n$, because $N_0+N_1=1$. This fact allows us to plot the
electronic relaxation rate $-dN_1/dt$ simply as a function of $N_0$.
The dependence of the relaxation rate on $N_0$ is shown in the Fig.\
\ref{tri}.
The plot is shown for three values of the quantum dot size, in which
the corresponding inter-level separation, or the detuning, is set up
to be such, that $E_1=\xi E_{LO}$, with $\xi$ being 0.7, 1 and 1.7.
The value of $\xi=1$ corresponds to the resonance between $E_1$ and
$E_{LO}$.   The relaxation rate is not constant as a function of
$N_0$. Also, the relaxation rate  depends on the magnitude of the
detuning $\xi$.  The relaxation rate changes sign at some values of
$N_0$, which are different from 0 and 1. Another  numerical result
says that  the dependence of the rate on the population is
monotoneous. We see in  Fig.
\ref{tri}  that at all three cases of the detuning $\xi$
the relaxation rate becomes zero at
the population,
 which is about between 0.1
and 0.2 at the lattice temperature of 10\,K. Realizing that the optical
phonon energy in GaAs is about 36.2 meV, the electronic inter-level
energy separations considered in this figure are of the same order of
magnitude. The value of $N_0$ in Fig.\ \ref{tri}, at which the rate is
zero and the relaxation stops, is therefore not expected to be simply
due to the equilibrium effect of the lattice temperature electronic
level population.

\begin{figure}[t]
\includegraphics[width=80mm]{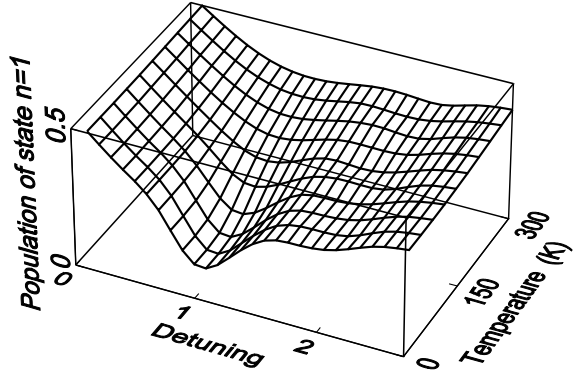}%
 \caption{The final population
  $N_{1f}$, at which the relaxation rate is zero,
   as a function of  detuning $\xi$ and  lattice temperature $T$.
  }
 \label{n1a}
\end{figure}

Fig.\ \ref{tri} shows that the final state of the electronic relaxation
from the state with the population  $N_1=1$, at low temperatures,
cannot be the state with $N_1=0$, but it should be another state of the
sytem, in which  the population of the ground state $n=0$ is different
from one. In other words, the electron  does not relax to the ground
state even in the limit of low temperatures. In Fig.\
\ref{n1a} we plot the final population $N_{1f}$ of the excited state,
which is defined as such a value of the population $N_1$, at which the
relaxation rate becomes zero. The quantity $N_{1f}$ is plotted as a
function of the detuning and lattice temperature.
We expect that due to the population dependence of the relaxation
rate,  shown in Fig.\
\ref{tri},  the population of the state $n=1$ will be always
either decreasing at $t>0$, if we start with $N_1=1$ at $t=0$, or
increasing, if we start with the population $N_1=0$ at $t=0$. The
population of the states will be developing towards  the limiting
population $N_{1f}$ plotted in Fig.\
\ref{n1a}.

 \begin{figure}[t]
\includegraphics[width=80mm]{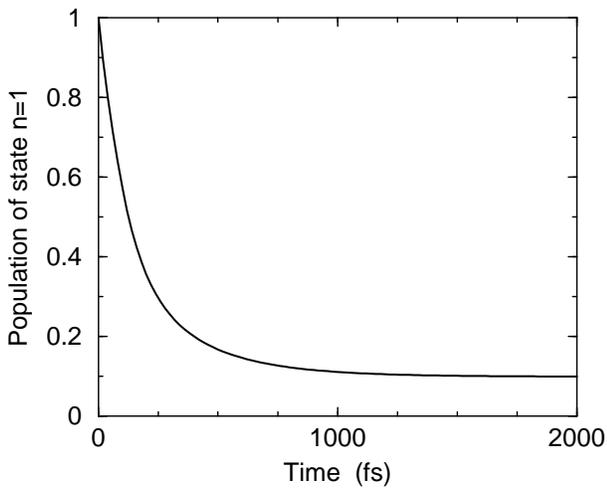}%
 \caption{Time dependence of population $N_1$
 of the higher-energy electronic state $n=1$ for the detuning $\xi$=1,
 ($E_1=E_{LO}$), at lattice temperature $T$=10 K.}
 \label{024}
\end{figure}

 \begin{figure}[t]
\includegraphics[width=80mm]{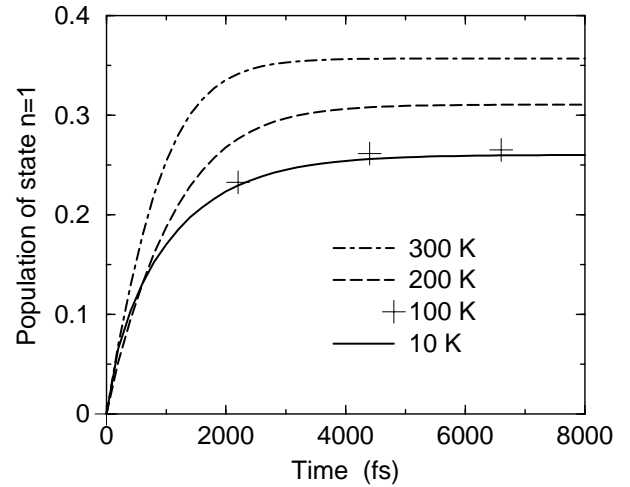}%
 \caption{Time evolution of the population $N_1$,
 of the higher-energy state $n=1$, at the detuning $\xi$=0.7.}
 \label{evol0.7}
\end{figure}

The final value $N_{1f}$ of $N_1$ depends both on lattice temperature
and on the detuning $\xi$. The quantity $N_{1f}$ appears to increase
with increasing the lattice temperature, which is expected. The
dependence of $N_{1f}$ on the detuning $\xi$ displays a certain
modulation. At the integer values of $\xi$, which correspond to the
resonance between the excitation energy $E_1$ and a multiple of the LO
phonon energy, the value of $N_{1f}$ shows a minimum, which is rather
pronounced especially in the limit of low temperature and near $\xi=1$.
At higher lattice temperatures the latter resonance structure appears
to weaken. The present two-level model shows that the limiting value
$N_{1f}$ is especially large, achieving the magnitude of about 0.4 to
0.5, when the detuning is either much larger, or much lower than 1.

Studying the time dependence of the electronic distribution in the
process of the electronic relaxation, let us first assume that the
electron is prepared in the state $n=1$ at $t=0$.  The numerical result
is displayed in Fig.\ \ref{024}. On the  time scale of picoseconds, the
population of the excited state decreases with time, reaching the
limiting value of the electronic population, in agreement with the data
shown in Fig.\
\ref{n1a}.

 \begin{figure}[t]
\includegraphics[width=80mm]{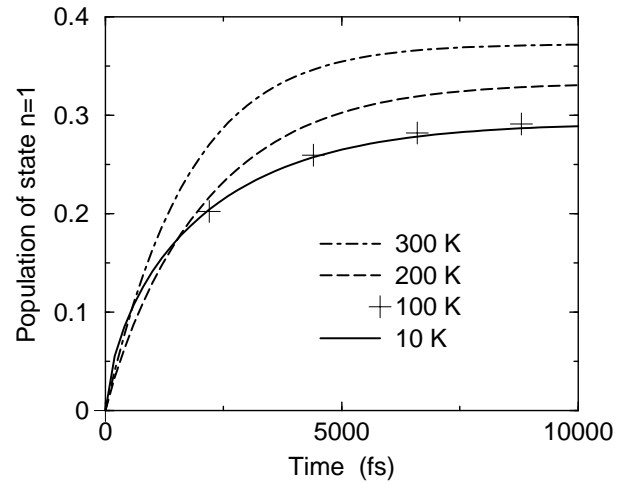}%
 \caption{Time evolution of the population $N_1$,
 of the higher-energy state $n=1$, at the detuning $\xi$=1.7.}
 \label{evol1.7}
\end{figure}
It is very interesting to observe  the time development of the
electronic distribution in the present two-level model in the
particular case when the electron is prepared in the ground state
$n=0$, namely when the population $N_1=0$ is prepared at time $t=0$. 
This innitial condition may correspond to the experiments, in which 
time resolved depopulation of the energy levels is measured, including 
the time dependence of the optical properties of quantum dots after 
exciting the quantum dots  to the lowest energy state of the 
electronic subsystem in a single quantum dot
\cite{quochi,urayama2001}. The time dependence of the
population $N_1$ corresponding to this initial condition is displayed
in Figs.\
\ref{evol0.7} and \ref{evol1.7}, in which this time dependence
is shown for two values of the detuning $\xi$. From these figures it 
is seen that on the time scale  of picoseconds the population of the 
electronic excited  state can increases to about 0.2 to 0.3, which 
means that the population of the lower energy state can decrease to 
0.8 to 0.7. This happens in the limit of low temperatures, at which 
the thermal population of the higher-energy states would be 
unexpected. From the difference between Figs.\ \ref{evol0.7} and 
\ref{evol1.7} we see that the characteristic  time of this population  
upconversion slightly varies with detuning and temperature. In 
agreement with the data shown in the Figs.\
\ref{n1a}, the results shown in Figs.\ \ref{evol0.7} and \ref{evol1.7}
demonstrate
 that at temperatures below
about 100 K the rapidity of the population changes does not depend
significantly on temperature.

\section{\label{}Discussion}

The present theoretical results can find a counterpart in the recent
experiments in which the time resolved response of quantum dots to
short laser pulse is studied \cite{urayama2001,quochi}.  The
experimental experience shows that when
 a quantum dot is excited with a short laser  pulse to the
 lowest-energy state, the subsequent relaxation process
  may lead to the appearence of the electronic population of electronic
  states with higher energy
\cite{quochi}. Another effect found in experiment is:  the system
of quantum dot, when prepared in an excited state, relaxes fast from
this state, but the relaxation ends without depopulating completely the
excited state \cite{urayama2001}. In these experiment the depopulation
effects do not show any dependence on the resonance between the
electronic inter-level energy separation and optical phonon energy.
These effects were recently interpreted to be a result of a
non-geminate excitation of the electrons and holes in quantum dots
\cite{urayama2001,urayama2002} or as a manifestation of multiphonon
scattering processes including acoustic phonons
\cite{quochi}.

Besides the interpretations of the origin of these depopulation
effects  given in the experimental papers
\cite{urayama2001,quochi,urayama2002},
we propose the mechanism studied in this
paper as another possible source of these effects. This mechanism
assumes the absence of the phonon bottleneck effect: The formula
(\ref{rate}) for the relaxation rate does not contain the energy
conservation delta function factor. It describes the electronic
transitions between two electronic states, which are bound to a cloud
of optical phonons. This coherent phonons cloud is induced thanks to
the electron-phonon coupling  by the virtual transitions of the
electron between the two electronic states. Because these multiphonon
states are not necessarily Fock states with well-defined number of LO
phonons, the energy conserving delta function does not appear in the
relaxation rate formula. The coherent multiphonon states can be seen
as  classical macroscopic oscillations of the optical phonon modes.
These macroscopic oscillations can be regarded as a source of an
effective electronic Hamiltonian depending explicitly on time, making
the system nonconservative.  In such a system we do not speak about
the energy conservation. This is the way how we can understand the
electronic relaxation in the case of nonresonant situation between the
inter-level energy separation and the phonon energy, both in the case
of the relaxation from the high energy level and in the case of the
relaxation of the electron from the low-energy level to the
high-energy one.

The coupling of the electron to the phonon system via multiple
scattering virtual transitions may cause a strong mixing of the two
electronic states causes. Because of the effective lack of the energy
conservation, the final state of the relaxation of the system  need
not to be identical with the low energy state, in the limit of zero
temperature.

Summing up, we have demonstrated the time dependence of electronic
relaxation in the  model  quantum dot with two levels and a single
electron. We demonstrate the incomplete depopulation of the excited
state and we also demonstrate the appearence of the population of the
excited state after the resonance excitation of the ground state. We
present the characteristic times of these processes. These theoretical
results are discussed in relation to the recent experiments and an
additional interpretation of these experiments is proposed.

\begin{acknowledgments}
This work was supported by grants IAA1010113, RN19982003014 and by the 
project AVOZ1-010-914.
\end{acknowledgments}

\bibliography{tecky}

\end{document}